\documentstyle[amstex,amssymb,preprint,aps]{revtex}

\begin{document}
\title{Recoil-Induced-Resonances in Nonlinear, Ground-State, Pump-Probe Spectroscopy}
\author{C. P. Search and P. R. Berman}
\address{Physics Department, University of Michigan, Ann Arbor, MI, 48109-1120}
\date{\today}
\maketitle
\pacs{32.80.-t, 42.65.-k, 32.70.Jz}

\begin{abstract}
A theory of pump-probe spectroscopy is developed in which optical fields
drive two-photon Raman transitions between ground states of an ensemble of
three-level atoms. Effects related to the recoil the atoms undergo as a
result of their interactions with the fields are fully accounted for in this
theory. The linear absorption coefficient of a weak probe field in the
presence of two pump fields of arbitrary strength is calculated. For
subrecoil cooled atoms, the spectrum consists of eight absorption lines and
eight emission lines. In the limit that $\chi _{1}\ll \chi _{2}$, where $%
\chi _{1}$ and $\chi _{2}$ are the Rabi frequencies of the two pump fields,
one recovers the absorption spectrum for a probe field interacting with an
effective two-level atom in the presence of a single pump field. However
when $\chi _{1}\gtrsim \chi _{2}$, new interference effects arise that allow
one to selectively turn on and off some of these recoil induced resonances.
\end{abstract}

\section{Introduction}

Recent advances in laser cooling, atom optics, and Bose-Einstein
condensation have underlined the role played by atomic recoil in atom-field
interactions. A measure of the importance of recoil effects is the recoil
frequency, $\omega _{\hbar {\bf k}}=\hbar k^{2}/2M,$ associated with the
absorption, emission or scattering of radiation of wavelength $\lambda =2\pi
/k$ by an atom of mass $M.$ Once this quantity becomes greater than or
comparable to decay rates or Doppler widths that characterize the spectral
response of atoms, recoil can lead to new features in absorption or emission
line shapes. One class of such phenomena has been termed {\em recoil-induced
resonances} (RIR) [1-8], which occur when a weak probe and strong pump field 
{\em simultaneously} drive a given atomic transition. Interesting in their
own right, the RIR have been used to determine the velocity distribution of
laser-cooled atoms \cite{RIR7}, as a probe of Bose-Einstein condensates \cite
{bragg1}\cite{bragg2}, and in a feedback mechanism in stochastic cooling 
\cite{raizen}. Related to the RIR is the so-called {\em collective atomic
recoil laser} (CARL), which operates on similar principles but in a somewhat
different parameter range \cite{carl}. Both the RIR and CARL represent new
diagnostic probes of cold-atom sysytems. Recently, Moore and Meystre \cite
{moore} proposed that CARL be used to entangle optical and matter fields, as
well as to entangle different modes of the condensate excited by the optical
fields. Our discussion is limited to situations in which the collective
effects associated with CARL can be neglected. In this paper, we combine RIR
with {\em ground state spectroscopy} \cite{ground-state} to obtain
qualitatively new features in the probe absorption spectrum.

The scheme we adopt is based on the model developed in \cite{ground-state}
involving Raman transitions, but for which all effects associated with
atomic recoil were ignored. In that work a new type of interference was
discovered, allowing one to selectively turn on and off certain lines in the
absorption-emission spectrum by controlling the ratio of the Rabi
frequencies of the two fields that comprise the two-photon pump field.
Interference in a dressed state basis occurs between pathways involving the
probe field and each of the two pump fields. Since the two pump fields
impart different recoil momenta to the atoms, it is not at all obvious that
the interference persists when recoil splittings are resolved. Part of the
motivation for our calculation is to examine this question. In addition, we
show that the interference persists even if the pump fields are in
quantized, Fock states.

The probe absorption spectrum consists of as many as eight
absorption-emission doublets which are fully resolvable if $\omega
_{k}>\gamma $, where $\gamma $ is some effective ground state lifetime. This
is in contrast to the RIR spectrum on dipole allowed optical transitions 
\cite{RIR1}, where at most one absorption emission doublet is resolvable if $%
\omega _{k}<\gamma _{e}$, where $\gamma _{e}$ is an {\em excited} state
decay rate. One might question the need to increase the number of recoil
doublets in the probe spectrum, since a single doublet can be used to probe
recoil effects. We show that the additional recoil structure reflects
interesting quantum dynamics of the combined atom-field system, as well as
providing some new applications.

In Sect. II, a model is developed for the interaction of the atoms with the
pump fields; dressed states of the atom plus pump fields are defined. In
Sect. III, the interaction with the probe field is introduced and the
dressed state picture is used to obtain the probe absorption spectrum in the
secular limit. In Sect. IV we discuss the results and possible applications.
Nonsecular contributions to the absorption coefficient are calculated in an
Appendix.

\section{Equations of Motion}

The atom-field system is indicated schematically in Fig.1. Ground state
levels $|1>$ and $|2>$ are pumped incoherently with rates $\Lambda _{1}({\bf %
p)}$ and $\Lambda _{2}({\bf p})$, respectively, and the population of both
states decay with rate $\gamma $. If states $|1>$ and $|2>$ represent stable
ground states of the atom, then the pumping rates and decay rate constitute
a simple model for atoms that enter and leave the interaction volume. The
ground to excited state transition frequencies are denoted by $\omega _{ej}$ 
$(j=1,2)$. The pump fields 1 and 2, which constitute the two-photon pump
field, are denoted by the coupling constants $g_{1}$ and $g_{2}$,
respectively. Pump field 1 couples only state $|1>$ and excited state $|e>$,
while pump 2 couples only states $|2>$ and $|e>$. The pump fields have
frequencies $\Omega _{1}$ and $\Omega _{2}$, and propagation vectors ${\bf k}%
_{1}$ and ${\bf k}_{2}$, respectively. In this section, equations are
derived for the atom-pump field interaction, neglecting the incoherent
pumping and decay of the ground state levels. In the following section, the
interaction of the atoms with the probe field, $E_{p}$, which couples only
states $|1>$ and $|e>,$ is incorporated into the calculation, as are the
incoherent pumping rates $\Lambda _{1}({\bf p)}$ and $\Lambda _{2}({\bf p})$%
, and the ground state decay rate $\gamma $.

In contrast to \cite{ground-state} but as in \cite{RIR3}, we use a quantized
description of the pump fields. If the pump field detunings, 
\begin{mathletters}
\begin{equation}
\Delta _{2}=\Omega _{2}-\omega _{e2}\approx \Delta _{1}=\Omega _{1}-\omega
_{e1}\equiv \Delta ,
\end{equation}
are sufficiently large such that $\left| \chi _{1,2}/\Delta \right| ^{2}\ll
1 $ and $\left| \gamma _{e}/\Delta \right| \ll 1$, where the $\chi _{1,2}$
are defined below and $\gamma _{e}$ is the excited state decay rate, it is
possible to adiabatically eliminate the excited state to arrive at an
effective Hamiltonian involving only states $|1>$ and $|2>$ which is of the
form 
\end{mathletters}
\begin{mathletters}
\begin{eqnarray}
H &=&H_{a}+H_{r}+H_{ar};  \label{1} \\
H_{a} &=&\sum_{{\bf p}}\left[ (\hbar \omega _{1}+\frac{{\bf p}^{2}}{_{2M}}%
)|1,{\bf p}><1,{\bf p}|+(\hbar \omega _{2}+\frac{{\bf p}^{2}}{_{2M}})|2,{\bf %
p}><2,{\bf p}|\right] ; \\
H_{r} &=&\hbar \Omega _{1}a_{1}^{\dagger }a_{1}+\hbar \Omega
_{2}a_{2}^{\dagger }a_{2}; \\
H_{ar} &=&\sum_{{\bf p}}\left[ \hbar \frac{g_{1}g_{2}^{\ast }e^{i{\bf k}%
_{12}\cdot {\bf R}}}{\Delta }|2,{\bf p}><1,{\bf p}|a_{2}^{\dagger
}a_{1}+\hbar \frac{g_{2}g_{1}^{\ast }e^{i{\bf k}_{21}\cdot {\bf R}}}{\Delta }%
|1,{\bf p}><2,{\bf p}|a_{1}^{\dagger }a_{2}\right] ,  \label{twophot}
\end{eqnarray}
where $H_{a}$ is the Hamiltonian for the atom in which the center-of-mass
momentum ${\bf p}$ has been quantized using periodic boundary conditions in
a volume $V$ (assuming that the atoms are free and not subject to some
trapping potential), $H_{r}$ is the free field Hamiltonian for the two pump
fields, $H_{ar}$ represents the interaction of the two-photon pump field
with an atom, in the rotating-wave approximation, and 
\end{mathletters}
\[
{\bf k}_{ij}={\bf k}_{i}-{\bf k}_{j}\text{. } 
\]
Bare states, $|j,{\bf p};n_{1},n_{2}>,$ are defined to be eigenstates of $%
H_{a}+H_{r}$, where $j=1,2$ labels the internal state of the atom and $n_{1}$
and $n_{2}$ are the number of photons in pump fields 1 and 2, respectively.
A term in the Hamiltonian corresponding to the light shifts of the ground
state levels, $H_{ls}=\sum_{{\bf p}}\left[ \hbar \frac{\left| g_{1}\right|
^{2}}{\Delta }|1,{\bf p}><1,{\bf p}|a_{1}^{\dagger }a_{1}+\hbar \frac{\left|
g_{2}\right| ^{2}}{\Delta }|2,{\bf p}><2,{\bf p}|a_{2}^{\dagger }a_{2}\right]
,$ has been omitted in Eq. (\ref{1}); such light shifts can be included by a
redefinition of the ground state frequencies, $\omega _{2}+\frac{\left|
g_{2}\right| ^{2}}{\Delta }\left\langle a_{2}^{\dagger }a_{2}\right\rangle
\rightarrow \omega _{2}$ and $\omega _{1}+\frac{\left| g_{1}\right| ^{2}}{%
\Delta }\left\langle a_{1}^{\dagger }a_{1}\right\rangle \rightarrow \omega
_{1.}$

The matrix elements of the operator $e^{i{\bf k}\cdot {\bf R}}$ in the
momentum-state basis are 
\begin{equation}
<{\bf p}|e^{i{\bf k}\cdot {\bf R}}|{\bf p}^{\prime }>=<{\bf p}|{\bf p}%
^{\prime }+\hbar {\bf k}>=\delta _{{\bf p},{\bf p}^{\prime }+\hbar {\bf k}}.
\end{equation}
This allows one to rewrite the interaction term as 
\begin{mathletters}
\begin{equation}
H_{ar}=\sum_{{\bf p}}\left[ \hbar \frac{g_{1}g_{2}^{\ast }}{\Delta }|2,{\bf p%
}+\hbar {\bf k}_{12}><1,{\bf p}|a_{2}^{\dagger }a_{1}+\hbar \frac{%
g_{2}g_{1}^{\ast }}{\Delta }|1,{\bf p}><2,{\bf p}+\hbar {\bf k}%
_{12}|a_{1}^{\dagger }a_{2}\right] .
\end{equation}
The Hamiltonian $H$ results in an infinite ladder of decoupled two state
manifolds $({\bf p},n_{1},n_{2})$ involving the states $|1,{\bf p}%
;n_{1},n_{2}>$ and $|2,{\bf p}+\hbar {\bf k}_{12};n_{1}-1,n_{2}+1>$. The
Hamiltonian for the manifold $({\bf p},n_{1},n_{2})$ is 
\end{mathletters}
\begin{equation}
H({\bf p},n_{1},n_{2})=\varepsilon ({\bf p},n_{1},n_{2}){\bf I}+\hbar \left( 
\begin{array}{cc}
-\tilde{\delta}({\bf p})/2 & G^{\ast } \\ 
G & \tilde{\delta}({\bf p})/2
\end{array}
\right)  \label{H'}
\end{equation}
where ${\bf I}$ is the identity matrix 
\[
G=\frac{\chi _{1}^{\ast }\chi _{2}}{\Delta }\equiv |G|e^{i\phi _{d}}, 
\]
\begin{equation}
\varepsilon ({\bf p},n_{1},n_{2})=\hbar (n_{1}\Omega _{1}+n_{2}\Omega _{2})+%
\frac{\hbar }{2}(\omega _{{\bf p}}+\omega _{{\bf p}+\hbar {\bf k}%
_{12}}+\omega _{1}+\omega _{2}+\Omega _{2}-\Omega _{1}),
\end{equation}
\begin{eqnarray*}
\tilde{\delta}({\bf p}) &=&\delta _{12}-\omega _{\hbar {\bf k}_{12}}-\frac{%
{\bf p}\cdot {\bf k}_{12}}{M}; \\
\delta _{12} &=&\Delta _{1}-\Delta _{2}=\left( \Omega _{1}-\Omega
_{2}\right) -\omega _{21},
\end{eqnarray*}
$\hbar \omega _{{\bf p}}=\frac{{\bf p}^{2}}{_{2M}}$, and $\chi _{2}=g_{2}%
\sqrt{n_{2}+1}$, $\chi _{1}=g_{1}\sqrt{n_{1}}.$

The dressed states are defined to be the eigenstates of the matrix in Eq.(%
\ref{H'}), with energies 
\begin{eqnarray*}
E_{B,A}({\bf p}) &=&\varepsilon ({\bf p},n_{1},n_{2})\pm \frac{\hbar \omega
_{AB}({\bf p})}{2}; \\
\omega _{AB}({\bf p}) &=&\sqrt{4|G|^{2}+\tilde{\delta}({\bf p})^{2}},
\end{eqnarray*}
and associated eigenkets \cite{convert} 
\begin{equation}
\left( 
\begin{array}{c}
|A_{0}> \\ 
|B_{0}>
\end{array}
\right) ={\bf T}^{\ast }({\bf p)}\left( 
\begin{array}{c}
|2,{\bf p}+\hbar {\bf k}_{12};n_{1}-1,n_{2}+1> \\ 
|1,{\bf p};n_{1},n_{2}>
\end{array}
\right)  \label{10}
\end{equation}
where 
\begin{equation}
{\bf T}({\bf p})=\left( 
\begin{array}{cc}
e^{i\phi _{d}/2}\cos \left[ \theta ({\bf p})\right] & -e^{-i\phi _{d}/2}\sin 
\left[ \theta ({\bf p})\right] \\ 
e^{i\phi _{d}/2}\sin \left[ \theta ({\bf p})\right] & e^{-i\phi _{d}/2}\cos 
\left[ \theta ({\bf p})\right]
\end{array}
\right)
\end{equation}
and 
\begin{mathletters}
\begin{equation}
\cos \left[ \theta ({\bf p})\right] =\left[ \frac{1}{2}\left( 1+\frac{\tilde{%
\delta}({\bf p})}{\omega _{AB}({\bf p})}\right) \right] ^{1/2}.
\end{equation}
The value of $\theta ({\bf p})$ is restricted to the range $0\leq \theta (%
{\bf p})\leq \pi /4$ for $\delta ({\bf p})\geq 0$ and $\pi /4\leq \theta (%
{\bf p})\leq \pi /2$ for $\tilde{\delta}({\bf p})\leq 0.$ For $\theta ({\bf p%
})\sim 0$ ($\tilde{\delta}({\bf p})>0$ and $\tilde{\delta}({\bf p})/|G|\gg 1$%
), $|A_{0}>\sim |2,{\bf p}+\hbar {\bf k}_{12};n_{1}-1,n_{2}+1>$ while for $%
\theta ({\bf p})\sim \pi /2$ ($\tilde{\delta}({\bf p})<0$ and $\left| \tilde{%
\delta}({\bf p})\right| /|G|\gg 1$), $|B_{0}>\sim |2,{\bf p}+\hbar {\bf k}%
_{12};n_{1}-1,n_{2}+1>$.

\section{Probe Field Absorption in the Secular Limit}

The effect of the probe field is to induce transitions between states in
different manifolds. As is customary in dressed atom approaches, the probe
is treated as a classical field, 
\end{mathletters}
\begin{equation}
{\bf E}({\bf R,}t)=\frac{1}{2}\widehat{{\bf \epsilon }}E_{p}e^{i({\bf k}%
_{p}\cdot {\bf R}-\Omega _{p}t)}+c.c.,
\end{equation}
where $\widehat{{\bf \epsilon }}$ is a unit polarization vector. For our
problem, however, this choice represents a {\em hybrid} approach, since
two-quantum processes involving the probe field and either of the pump
fields mix classical and quantized fields. Although the probe field is
treated classically, its effect on the {\em momentum} of the states must be
accounted for explicitly. If the probe field detuning on the $1\rightarrow e$
transition is sufficiently large to be consistent with the adiabatic
elimination of the excited state, all transitions involving the probe occur
via two-quantum transitions involving the probe field and either of the pump
fields. In the bare state basis, starting from the $({\bf p},n_{1},n_{2})$
manifold, probe field absorption corresponds to transitions $|1,{\bf p}%
;n_{1},n_{2}>\rightarrow |1,{\bf p}+\hbar {\bf k}_{p1};n_{1}+1,n_{2}>$ {\it %
or }$|1,{\bf p};n_{1},n_{2}>\rightarrow |2,{\bf p}+\hbar {\bf k}%
_{p2};n_{1},n_{2}+1>$, where the second photon is emitted into either the
pump 1 or pump 2 modes, respectively. Similarly, probe gain corresponds to
transitions $|1,{\bf p};n_{1},n_{2}>\rightarrow $ $|1,{\bf p}-\hbar {\bf k}%
_{p1};n_{1}-1,n_{2}>$ {\it or} $|2,{\bf p}+\hbar {\bf k}%
_{12};n_{1}-1,n_{2}+1>\rightarrow |1,{\bf p}-\hbar {\bf k}%
_{p1};n_{1}-1,n_{2}>$. This is illustrated in Fig. 2(a).

This picture of probe field absorption or emission allows us to reach an
important conclusion concerning interference between pathways involving both
pump fields. The two absorption processes shown in Fig. 2(a) involve
different final states and do not interfere, reinforcing the possibility
mentioned in the Introduction that interference may be suppressed when
recoil is taken into account. However, both final states belong to the {\em %
same} manifold - the $({\bf p}+\hbar {\bf k}_{p1},n_{1}+1,n_{2})$ manifold.
As such, when these states are dressed by the two-photon pump field, each
state in the final state manifold will be coupled to the initial state by 
{\em two}{\bf \ }separate pathways involving the probe field and each of the
pump fields. This implies that terms in the probe absorption depending on
the {\em simultaneous} presence of both pump fields will exhibit
interference effects. In the case of probe gain, it is the {\em initial}
states that differ, but the overall conclusion remains unchanged.

The dressed state approach provides a convenient and relatively easy method
for obtaining the probe absorption spectrum in the secular limit, where the
frequency separation of the dressed states in a given doublet is much larger
than the ground state decay rate $\gamma $. [It is assumed from this point
onward that $\gamma _{e}\left( \chi _{1,2}/\Delta \right) ^{2}\ll \gamma ,$
implying that the dressed states decay with rate $\gamma $ \cite{RIR3}]. One
need calculate only the dressed state energies and the transition matrix
elements to obtain the spectrum. A more detailed treatment of the problem,
allowing one to calculate non-secular contributions, is presented in
Appendix B.

It is straightforward to generalize the dressed states defined in the
previous section to include the two manifolds coupled to the initial
manifold by the probe. The 0, 1, and 2 manifolds refer to $({\bf p}%
,n_{1},n_{2})=\left\{ |1,{\bf p};n_{1},n_{2}>,|2,{\bf p}+\hbar {\bf k}%
_{12};n_{1}-1,n_{2}+1>\right\} $, $({\bf p}+\hbar {\bf k}%
_{p1},n_{1}+1,n_{2})=\left\{ |1,{\bf p}+\hbar {\bf k}_{p1};n_{1}+1,n_{2}>,|2,%
{\bf p}+\hbar ({\bf k}_{12}+{\bf k}_{p1});n_{1},n_{2}+1>\right\} ,$ and $(%
{\bf p}-\hbar {\bf k}_{p1},n_{1}-1,n_{2})=\left\{ |1,{\bf p}-\hbar {\bf k}%
_{p1};n_{1}-1,n_{2}>,|2,{\bf p}+\hbar ({\bf k}_{12}-{\bf k}%
_{p1});n_{1}-2,n_{2}+1>\right\} $, respectively. Taking the central energy
of the initial manifold $({\bf p},n_{1},n_{2})$ arbitrarily equal to zero,
one finds that the dressed state energies are given by 
\begin{eqnarray}
E_{A,B}^{(0)} &=&\pm \frac{1}{2}\hbar \omega _{AB}^{(0)}({\bf p});  \nonumber
\\
E_{A,B}^{(1)} &=&\hbar \omega _{10}({\bf p})\pm \frac{1}{2}\hbar \omega
_{AB}^{(1)}({\bf p});  \label{Edress} \\
E_{A,B}^{(2)} &=&\hbar \omega _{20}({\bf p})\pm \frac{1}{2}\hbar \omega
_{AB}^{(2)}({\bf p}),  \nonumber
\end{eqnarray}
where 
\begin{equation}
\omega _{AB}^{(i)}({\bf p})=\sqrt{4|G|^{2}+\delta _{i}({\bf p})^{2}},
\label{dress_split}
\end{equation}
\begin{eqnarray}
\delta _{0}({\bf p}) &=&\tilde{\delta}({\bf p})=\delta _{12}-\omega _{\hbar 
{\bf k}_{12}}-\frac{{\bf p}\cdot {\bf k}_{12}}{M};  \nonumber \\
\delta _{1}({\bf p}) &=&\tilde{\delta}({\bf p}+\hbar {\bf k}_{p1})=\delta
_{12}-\omega _{\hbar {\bf k}_{12}}-\frac{({\bf p}+\hbar {\bf k}_{p1})\cdot 
{\bf k}_{12}}{M};  \label{man_detun} \\
\delta _{2}({\bf p}) &=&\tilde{\delta}({\bf p}-\hbar {\bf k}_{p1})=\delta
_{12}-\omega _{\hbar {\bf k}_{12}}-\frac{({\bf p}-\hbar {\bf k}_{p1})\cdot 
{\bf k}_{12}}{M},  \nonumber
\end{eqnarray}
\begin{mathletters}
\begin{eqnarray}
\hbar \omega _{10}({\bf p}) &=&\varepsilon ({\bf p}+\hbar {\bf k}%
_{p1},n_{1}+1,n_{2})-\varepsilon ({\bf p},n_{1},n_{2})=\hbar \left( \Omega
_{1}+\omega _{\hbar {\bf k}_{p1}}+\frac{{\bf k}_{p1}\cdot {\bf p}}{M}+\frac{%
\hbar {\bf k}_{p1}\cdot {\bf k}_{12}}{2M}\right) ; \\
\hbar \omega _{20}({\bf p}) &=&\varepsilon ({\bf p}-\hbar {\bf k}%
_{p1},n_{1}-1,n_{2})-\varepsilon ({\bf p},n_{1},n_{2})=\hbar \left( -\Omega
_{1}+\omega _{\hbar {\bf k}_{p1}}-\frac{{\bf k}_{p1}\cdot {\bf p}}{M}-\frac{%
\hbar {\bf k}_{p1}\cdot {\bf k}_{12}}{2M}\right) ,
\end{eqnarray}
and it has also been assumed that $n_{1},n_{2}\gg 1$ such that $%
G^{(0)}=g_{1}^{\ast }g_{2}\sqrt{n_{1}\left( n_{2}+1\right) }/\Delta \approx
G^{(1)}=g_{1}^{\ast }g_{2}\sqrt{\left( n_{1}+1\right) \left( n_{2}+1\right) }%
/\Delta \approx G^{(2)}=g_{1}^{\ast }g_{2}\sqrt{\left( n_{1}-1\right) \left(
n_{2}+1\right) }/\Delta \equiv G.$ The dressed state angles are given by 
\end{mathletters}
\begin{mathletters}
\begin{equation}
\cos \left[ \theta _{i}({\bf p})\right] =\left[ \frac{1}{2}\left( 1+\frac{%
\delta _{i}({\bf p})}{\omega _{AB}^{\left( i\right) }({\bf p})}\right) %
\right] ^{1/2}
\end{equation}
and dressed state kets are defined by 
\end{mathletters}
\begin{mathletters}
\begin{equation}
\left( 
\begin{array}{c}
|A_{i}> \\ 
|B_{i}>
\end{array}
\right) ={\bf T}_{i}^{\ast }({\bf p})\left( 
\begin{array}{c}
|2,{\bf p}+\hbar \left[ {\bf k}_{12}{\bf -}(-1)^{i}\left( 1-\delta
_{i,0}\right) {\bf k}_{p1}\right] ;n_{1}-1-(-1)^{i}\left( 1-\delta
_{i,0}\right) ,n_{2}+1> \\ 
|1,{\bf p-}(-1)^{i}\left( 1-\delta _{i,0}\right) \hbar {\bf k}%
_{p1};n_{1}-(-1)^{i}\left( 1-\delta _{i,0}\right) ,n_{2}>
\end{array}
\right)  \label{bare2dressed}
\end{equation}
where 
\end{mathletters}
\begin{equation}
{\bf T}_{i}({\bf p})=\left( 
\begin{array}{cc}
e^{i\phi _{d}/2}\cos \left[ \theta _{i}({\bf p})\right] & -e^{-i\phi
_{d}/2}\sin \left[ \theta _{i}({\bf p})\right] \\ 
e^{i\phi _{d}/2}\sin \left[ \theta _{i}({\bf p})\right] & e^{-i\phi
_{d}/2}\cos \left[ \theta _{i}({\bf p})\right]
\end{array}
\right) .  \label{T}
\end{equation}

The absorption coefficient, $\alpha $, is proportional to the rate at which
energy is absorbed from ($\alpha >0$) or emitted into ($\alpha <0$) the
probe field. Absorption corresponds to transitions from initial dressed
states $I=A_{0},B_{0}$ to final state dressed states $J=A_{1},B_{1}$, while
emission corresponds to transitions from initial dressed states $%
I=A_{0},B_{0}$ to final state dressed states $J=A_{2},B_{2}$. For a given
transition, the contribution to the absorption coefficient is proportional
to $NV\hbar \Omega _{p}\gamma \rho _{JJ}/|E_{p}|^{2}$, where $N$ is the atom
density and $\rho _{JJ}$ is the steady state population in state $J$ owing
to the $I\rightarrow J$ transition. The final state population $\rho _{JJ}(%
{\bf p)}$ is equal to $\left[ \Lambda _{I}({\bf p})/\gamma \right]
|<J|V_{p}|I>|^{2}L_{IJ}$, where $\Lambda _{I}({\bf p})$ is the pumping rate
for {\em initial} dressed state $I$ \cite{pumping}$,$ $<J|V_{p}|I>$ is a
matrix element for the $I-J$ transition and $L_{IJ}$ is a Lorentzian having
width $\gamma $, centered at the $I$-$J$ transition frequency$.$ The
transition frequencies $\Delta _{IJ}({\bf p})$ may be directly read from
Fig. 2(b) or obtained from Eq. (\ref{Edress}). Consequently, it is necessary
to specify $\Lambda _{I}({\bf p})$ and to calculate the transition matrix
elements $<J|V_{p}|I>$ in order to arrive at an expression for the probe
absorption.

In the bare state representation, the pumping matrix for the intitial state
manifold is taken to be of the form 
\[
{\bf \Lambda }({\bf p)}=\left( 
\begin{array}{cc}
\dot{\rho}_{11}({\bf p)} & 0 \\ 
0 & \dot{\rho}_{22}({\bf p+}\hbar {\bf k}_{12}{\bf )}
\end{array}
\right) _{pump}=\left( 
\begin{array}{cc}
\Lambda _{1}({\bf p)} & 0 \\ 
0 & \Lambda _{2}({\bf p+\hbar k_{12})}
\end{array}
\right) , 
\]
which, when converted to the dressed state basis becomes 
\begin{mathletters}
\label{pump}
\begin{eqnarray}
{\bf \Lambda }_{d}({\bf p}) &=&\left( 
\begin{array}{cc}
\Lambda _{A}({\bf p}) & \Lambda _{AB}({\bf p}) \\ 
\Lambda _{AB}({\bf p}) & \Lambda _{B}({\bf p})
\end{array}
\right) \\
&=&\left( 
\begin{array}{cc}
\Lambda _{2}({\bf p}+\hbar {\bf k}_{12})\cos ^{2}\left[ \theta _{0}({\bf p)}%
\right] +\Lambda _{1}({\bf p})\sin ^{2}\left[ \theta _{0}({\bf p)}\right] & 
\frac{1}{2}\left[ \Lambda _{2}({\bf p}+\hbar {\bf k}_{12})-\Lambda _{1}({\bf %
p})\right] \sin \left[ 2\theta _{0}({\bf p)}\right] \\ 
\frac{1}{2}\left[ \Lambda _{2}({\bf p}+\hbar {\bf k}_{12})-\Lambda _{1}({\bf %
p})\right] \sin \left[ 2\theta _{0}({\bf p)}\right] & \Lambda _{1}({\bf p}%
)\cos ^{2}\left[ \theta _{0}({\bf p)}\right] +\Lambda _{2}({\bf p}+\hbar 
{\bf k}_{12})\sin ^{2}\left[ \theta _{0}({\bf p)}\right]
\end{array}
\right) ;
\end{eqnarray}
In the secular limit, the off-diagonal pumping terms can be neglected since
they give rise to terms of order $\gamma /\omega _{AB}^{(i)}({\bf p})\ll 1$.
Moreover, for the present we will set $\Lambda _{2}({\bf p}+\hbar {\bf k}%
_{12}{\bf )=}0$ and generalize the results to nonvanishing $\Lambda _{2}(%
{\bf p}+\hbar {\bf k}_{12}{\bf )}$ in the next section. Thus, we take a
pumping matrix of the form 
\end{mathletters}
\[
{\bf \Lambda }_{d}\left( {\bf p}\right) =\left( 
\begin{array}{cc}
\Lambda _{1}\left( {\bf p}\right) \sin ^{2}\left[ \theta _{0}\left( {\bf p}%
\right) \right] & 0 \\ 
0 & \Lambda _{1}\left( {\bf p}\right) \cos ^{2}\left[ \theta _{0}\left( {\bf %
p}\right) \right]
\end{array}
\right) . 
\]

We now turn our attention to the transition matrix element $<J|V_{p}|I>.$ As
an example consider $<A_{1}|V_{p}|A_{0}>$, which is the amplitude for the
transition $|A_{0}>\rightarrow |A_{1}>$ involving probe absorption and pump
1 or pump 2 emission. This transition is illustrated in Fig. 2(c). The probe
couples only to the $|1,{\bf p};n_{1},n_{2}>$ part of $|A_{0}>,$ leading to
a factor $-e^{i\phi _{d}/2}\sin \left[ \theta _{0}({\bf p})\right] $. The
absorption of the probe is followed by emission into pump 2 taking the atom
to the $|2,{\bf p}+\hbar ({\bf k}_{12}+{\bf k}_{p1});n_{1},n_{2}+1>$
component of $|A_{1}>$ and leading to a factor $e^{-i\phi _{d}/2}\cos \left[
\theta _{1}({\bf p})\right] $ or emission into pump 1 taking the atom to the
the $|1,{\bf p}+\hbar {\bf k}_{p1};n_{1}+1,n_{2}>$ component and leading to
a factor $-e^{i\phi _{d}/2}\sin \left[ \theta _{1}({\bf p})\right] $. The
coupling strengths for these two, two-photon transitions are $G_{1}^{\ast }$
and $G_{2}^{\ast }$, respectively, where 
\begin{mathletters}
\label{pump}
\begin{equation}
G_{1}=\frac{\chi _{p}^{\ast }\chi _{1}}{\Delta };\text{ \ \ }G_{2}=\frac{%
\chi _{p}^{\ast }\chi _{2}}{\Delta }\text{\ }.
\end{equation}
The two processes add coherently and one finds 
\end{mathletters}
\[
<A_{1}|V|A_{0}>=\hbar \left\{ G_{2}^{\ast }e^{i\phi _{d}}\cos \left[ \theta
_{1}({\bf p})\right] -G_{1}^{\ast }\sin \left[ \theta _{1}({\bf p})\right]
\right\} \left\{ -\sin \left[ \theta _{0}({\bf p})\right] \right\}
e^{-i\Omega _{p}t}. 
\]
Other matrix elements are calculated in a similar manner. For probe gain,
pump fields 1 and 2 couple to the $|1,{\bf p};n_{1},n_{2}>$ and $|2,{\bf p}%
+\hbar {\bf k}_{12};n_{1}-1,n_{2}+1>$ components of the dressed states in
the 0 manifold, respectively, while the probe field couples to the $|1,{\bf p%
}-\hbar {\bf k}_{p1};n_{1}-1,n_{2}>$ component of the dressed states in the
2 manifold. Explicit expressions for the matrix elements are given in
Appendix A.

Combining all transitions and summing over {\bf p}, one finds an absorption
coefficient proportional to 
\begin{equation}
\alpha \varpropto \frac{\gamma ^{2}}{|\chi _{p}|^{2}}\sum_{{\bf p}}{\bf \,}%
\sum_{I=\{A_{o},B_{o}\}}\frac{\Lambda _{I}({\bf p})}{\gamma }\left(
\sum_{J=\{A_{1},B_{1}\}}\frac{|<J|V_{p}|I>|^{2}}{\left[ \Omega _{p}-\Delta
_{IJ}({\bf p})\right] ^{2}+\gamma ^{2}}-\sum_{J=\{A_{2},B_{2}\}}\frac{%
|<J|V_{p}|I>|^{2}}{\left[ \Omega _{p}-\Delta _{IJ}({\bf p})\right]
^{2}+\gamma ^{2}}\right) 
\end{equation}
For a sub-recoiled cooled vapor, we can set $\Lambda _{1}({\bf p})=\Lambda
_{1}\delta _{{\bf p},{\bf 0}}$, such that $\Lambda _{A}({\bf p})=\Lambda
_{A}\delta _{{\bf p},{\bf 0}}$ and $\Lambda _{B}({\bf p})=\Lambda _{B}\delta
_{{\bf p},{\bf 0}}$, where 
\[
\Lambda _{A}=\Lambda _{1}\sin ^{2}\theta _{0}\text{; \ \ \ \ \ }\Lambda
_{B}=\Lambda _{1}\cos ^{2}\theta _{0},\text{\ }
\]
$\theta _{i}\equiv \theta _{i}({\bf 0})$ [and, for future reference, $\Delta
_{IJ}\equiv \Delta _{IJ}({\bf 0})$, $\omega _{AB}^{\left( i\right) }\equiv
\omega _{AB}^{\left( i\right) }({\bf 0})$, etc.] and $\delta _{{\bf p},{\bf 0%
}}$ is a kronecker delta. In this limit, one finds the absorption
coefficient in the secular limit to be

\begin{mathletters}
\label{abs}
\begin{eqnarray}
\left( \frac{\alpha }{\alpha _{0}}\right) _{\sec } &=&\frac{|G|}{|\Delta |}%
\left[ \left( \psi \eta \sin \theta _{1}-\frac{1}{\eta }\cos \theta
_{1}\right) ^{2}\left( \frac{\Lambda _{A}}{\gamma }\sin ^{2}\theta
_{0}L_{A_{0}A_{1}}(\Delta ^{\prime })+\frac{\Lambda _{B}}{\gamma }\cos
^{2}\theta _{0}L_{B_{0}A_{1}}(\Delta ^{\prime })\right) \right.  \nonumber \\
&&+\left( \psi \eta \cos \theta _{1}+\frac{1}{\eta }\sin \theta _{1}\right)
^{2}\left( \frac{\Lambda _{A}}{\gamma }\sin ^{2}\theta
_{0}L_{A_{0}B_{1}}(\Delta ^{\prime })+\frac{\Lambda _{B}}{\gamma }\cos
^{2}\theta _{0}L_{B_{0}B_{1}}(\Delta ^{\prime })\right)  \nonumber \\
&&-\left( \psi \eta \sin \theta _{0}-\frac{1}{\eta }\cos \theta _{0}\right)
^{2}\frac{\Lambda _{A}}{\gamma }\left( \sin ^{2}\theta
_{2}L_{A_{0}A_{2}}(\Delta ^{\prime })+\cos ^{2}\theta
_{2}L_{A_{0}B_{2}}(\Delta ^{\prime })\right)  \nonumber \\
&&-\left. \left( \psi \eta \cos \theta _{0}+\frac{1}{\eta }\sin \theta
_{0}\right) ^{2}\frac{\Lambda _{B}}{\gamma }\left( \sin ^{2}\theta
_{2}L_{B_{0}A_{2}}(\Delta ^{\prime })+\cos ^{2}\theta
_{2}L_{B_{0}B_{2}}(\Delta ^{\prime })\right) \right] ;  \label{absorb2} \\
L_{A_{0}A_{1}}(\Delta ^{\prime }) &=&\frac{\gamma ^{2}}{\left( \Delta
^{\prime }-\omega _{\hbar {\bf k}_{p1}}-\frac{1}{2}(\omega
_{AB}^{(0)}-\omega _{AB}^{(1)})\right) ^{2}+\gamma ^{2}}; \\
L_{B_{0}A_{1}}(\Delta ^{\prime }) &=&\frac{\gamma ^{2}}{\left( \Delta
^{\prime }-\omega _{\hbar {\bf k}_{p1}}+\frac{1}{2}(\omega
_{AB}^{(0)}+\omega _{AB}^{(1)})\right) ^{2}+\gamma ^{2}}; \\
L_{A_{0}B_{1}}(\Delta ^{\prime }) &=&\frac{\gamma ^{2}}{\left( \Delta
^{\prime }-\omega _{\hbar {\bf k}_{p1}}-\frac{1}{2}(\omega
_{AB}^{(0)}+\omega _{AB}^{(1)})\right) ^{2}+\gamma ^{2}}; \\
L_{B_{0}B_{1}}(\Delta ^{\prime }) &=&\frac{\gamma ^{2}}{\left( \Delta
^{\prime }-\omega _{\hbar {\bf k}_{p1}}+\frac{1}{2}(\omega
_{AB}^{(0)}-\omega _{AB}^{(1)})\right) ^{2}+\gamma ^{2}}; \\
L_{A_{0}A_{2}}(\Delta ^{\prime }) &=&\frac{\gamma ^{2}}{\left( \Delta
^{\prime }+\omega _{\hbar {\bf k}_{p1}}+\frac{1}{2}(\omega
_{AB}^{(0)}-\omega _{AB}^{(2)})\right) ^{2}+\gamma ^{2}}; \\
L_{A_{0}B_{2}}(\Delta ^{\prime }) &=&\frac{\gamma ^{2}}{\left( \Delta
^{\prime }+\omega _{\hbar {\bf k}_{p1}}+\frac{1}{2}(\omega
_{AB}^{(0)}+\omega _{AB}^{(2)})\right) ^{2}+\gamma ^{2}}; \\
L_{B_{0}A_{2}}(\Delta ^{\prime }) &=&\frac{\gamma ^{2}}{\left( \Delta
^{\prime }+\omega _{\hbar {\bf k}_{p1}}-\frac{1}{2}(\omega
_{AB}^{(0)}+\omega _{AB}^{(2)})\right) ^{2}+\gamma ^{2}}; \\
L_{B_{0}B_{2}}(\Delta ^{\prime }) &=&\frac{\gamma ^{2}}{\left( \Delta
^{\prime }+\omega _{\hbar {\bf k}_{p1}}-\frac{1}{2}(\omega
_{AB}^{(0)}-\omega _{AB}^{(2)})\right) ^{2}+\gamma ^{2}};
\end{eqnarray}
where 
\end{mathletters}
\[
\psi =\Delta /|\Delta | 
\]
is the sign of the detuning for each of the three fields, 
\[
\eta =\sqrt{|\chi _{1}|/|\chi _{2}|}, 
\]
\begin{eqnarray*}
\Delta ^{\prime } &=&\delta _{p1}-\frac{\hbar {\bf k}_{p1}\cdot {\bf k}_{12}%
}{2M}, \\
\delta _{p1} &=&\Omega _{p}-\Omega _{1};\text{ \ \ \ }\delta _{p2}=\Omega
_{p}-\Omega _{2}-\omega _{21},
\end{eqnarray*}
\[
\alpha _{0}=\frac{k_{p}Nd_{1e}^{2}}{2\hbar \epsilon _{0}\gamma }, 
\]
and $d_{1e}$ is a bare state dipole moment matrix element for the $\left|
1\right\rangle \rightarrow \left| e\right\rangle $ transition.

The first four resonances in $\left( \frac{\alpha }{\alpha _{0}}\right)
_{\sec }$ correspond to probe absorption while the last four correspond to
probe gain. The spectrum is shown in Fig. 3. The line widths of all the
resonances equal $\gamma $; consequently, the recoil induced resonances in
ground state spectroscopy are fully resolved if $\omega _{k}>\gamma $. The
secular contribution does not vanish in the limit that $\left| \delta ({\bf p%
}={\bf 0})/G\right| \ll 1$, even though $\Lambda _{A}=\Lambda _{B}$ in this
limit. As long as the recoil frequency is larger than $\gamma $ and the
Doppler width associated with the two-photon pump transition, the absorption
and emission contributions to the probe response do not cancel one another.

The most significant feature of $\left( \frac{\alpha }{\alpha _{0}}\right)
_{\sec }$ is that the line strengths involve factors such as $\left( \psi
\eta \sin \theta _{1}-\frac{1}{\eta }\cos \theta _{1}\right) ^{2}$ which
allows one to manipulate the strength of the lines by controlling the sign
of the field detuning and the ratio of the pump field amplitudes. These
factors are an indication of interference between the two ''two-photon
probe'' fields which can both lead to absorption or gain in $E_{p}$. Because
the two lines in the doublets have different strengths, one can adjust $\psi 
$ and $\eta $ to turn off one of the lines. For example, the absorption
doublet, $L_{A_{0}A_{1}}$ and $L_{B_{0}B_{1}},$ consists of the lines at $%
\Delta ^{\prime }=\omega _{\hbar {\bf k}_{p1}}+\frac{1}{2}(\omega
_{AB}^{(0)}-\omega _{AB}^{(1)})$ and $\omega _{\hbar {\bf k}_{p1}}-\frac{1}{2%
}(\omega _{AB}^{(0)}-\omega _{AB}^{(1)})$ with strengths $\sim \left( \psi
\eta \sin \theta _{1}-\frac{1}{\eta }\cos \theta _{1}\right) ^{2}$ and $\sim
\left( \psi \eta \cos \theta _{1}+\frac{1}{\eta }\sin \theta _{1}\right)
^{2} $, respectively. Choosing $\psi =+1$ and $\eta ^{2}=\cot \theta _{1}$
turns ''off'' the first line while $\psi =-1$ and $\eta ^{2}=\tan \theta _{1}
$ turns ''off'' the second line. This is shown in Fig. 4. When $|G|$ and $%
\delta _{12}$ are much larger than any of the recoil terms, $\theta
_{0}\approx \theta _{1}\approx \theta _{2}$ and the emission lines are also
turned ''off''. Consequently, by choosing $\psi =+1$ and $\eta ^{2}=\cot
\theta _{1}$ to turn off the $L_{A_{0}A_{1}}$ and $L_{B_{0}A_{1}}$
absorption lines, the $L_{A_{0}A_{2}}$ and $L_{A_{0}B_{2}}$ emission lines
are also turned off.

A particularly interesting case occurs when ${\bf k}_{1}\approx {\bf k}_{2}$
so that $\omega _{AB}^{(0)}=\omega _{AB}^{(1)}=\omega _{AB}^{(2)}$ and $%
\theta _{0}=\theta _{1}=\theta _{2}$. This would correspond to a two-photon
pump field which imparts no momentum to the atoms so that the recoil
splitting in the absorption spectrum can be attributed solely to the recoil
due to the probe field acting with either of the pump fields, ${\bf k}_{p2}=%
{\bf k}_{p1}.$ In this case, the line $L_{A_{0}A_{1}}\left( \Delta ^{\prime
}\right) $ is degenerate with $L_{B_{0}B_{1}}\left( \Delta ^{\prime }\right) 
$ and $L_{A_{0}A_{2}}\left( \Delta ^{\prime }\right) $ is degenerate with $%
L_{B_{0}B_{2}}\left( \Delta ^{\prime }\right) .$ Consequently, the spectrum
consists of three absorption-emission doublets centered at $\Delta ^{\prime
}=0,+\omega _{AB}^{(0)},-\omega _{AB}^{(0)}.$ Moreover, the lines within
each doublet are split by $2\omega _{\hbar {\bf k}_{p1}}$ which is {\it %
independent} of the strength or detuning of the pump fields.

When the effects of atomic recoil are neglected by setting all recoil
momenta to zero in $\left( \frac{\alpha }{\alpha _{0}}\right) $, one obtains
the same absorption spectrum given in \cite{ground-state}. In the limit that 
$G_{1}=0,$ one recovers a simple, recoil shifted Raman spectrum. In the
limit that $G_{2}=0,$ one recovers the central, secular components of the
pump-probe spectrum associated with a {\em single}, two-level optical{\bf \ }%
transition \cite{RIR3}. In the limit that $\eta \ll 1$, while $G/\gamma \gg
1 $ remains constant, the absorption spectrum mirrors that for the
pump-probe spectrum associated with a two-level optical{\bf \ }transition 
\cite{RIR3}$.$

\section{Discussion}

For a subrecoil cooled vapor, pumping to state 2 at a rate $\Lambda
_{2}\left( {\bf p}+\hbar {\bf k}_{12}\right) =\Lambda _{2}\delta _{{\bf p}%
+\hbar {\bf k}_{12},{\bf 0}}$ doubles the number of absorption and emission
lines in the probe spectrum, but does not result in any qualitatively new
features. There will be an additional contribution to Eq. (\ref{abs}) in
which $\theta _{i}\equiv \theta _{i}({\bf 0})$ is replaced by $\bar{\theta}%
_{i}\equiv \theta _{i}({\bf -}\hbar {\bf k}_{12}),$ $\Delta _{IJ}$ by $\bar{%
\Delta}_{IJ}\equiv \Delta _{IJ}({\bf -}\hbar {\bf k}_{12}),$ $\omega
_{AB}^{i}$ by $\bar{\omega}_{AB}^{i}\equiv \omega _{AB}^{i}({\bf -}\hbar 
{\bf k}_{12}),\Lambda _{A}$ by $\Lambda _{2}\cos ^{2}\bar{\theta}_{0}$ and $%
\Lambda _{B}$ by $\Lambda _{2}\sin ^{2}\bar{\theta}_{0}.$ The absorption
coefficient contains sixteen lines in all. The eight new spectral components
display the same properties as the original eight but are displaced by an
amount $\sim \frac{\hbar {\bf k}_{p1}\cdot {\bf k}_{12}}{M}.$ Figure 5 shows
the secular absorption spectrum with all sixteen components when $\Lambda
_{2}=\Lambda _{1}$. In the absence of recoil, the secular absorption
coefficient vanishes when $\Lambda _{2}=\Lambda _{1}$ \cite{ground-state}.
However, when recoil is included, the absorption and emission contributions
to the probe response do not cancel one another when $\Lambda _{2}=\Lambda
_{1}$, provided the recoil frequency is larger than $\gamma $ and the
Doppler width associated with the two-photon pump transition.

In Ref. \cite{ground-state} interference effects similar to those in Eq. (%
\ref{absorb2}) were found. We have shown that the interference is
independent of the field statistics and persisits even when recoil induced
resonances are resolved.

The RIR offer several possibilities for applications. The existence of a
number of tunable, well-resolved, gain peaks allows one to envision
experiments in which lasing occurs at one or more of the gain positions.
These peaks could be adjusted to coincide with modes of a ring cavity, for
example. By sweeping the ratio of the two pump field intensities, one has a
mechanism for modifying the probe gain or absorption, to the point of total
suppression. The central frequency of the absorption-emission doublets can
be controlled via pump field strength and detuning. For parallel pump
fields, the frequency separation of the absorption emission doublets can be
as large as 8$\omega _{k}$ and is independent of pump field strength and
detuning. With line widths approaching 1 Hz or less, the probe spectrum can
be used to measure the recoil frequency to a precision of order $10^{-7}$;
this precision can be increased if the two-photon pump field is replaced by
a {\em pair} of counterpropagating, two-photon pump fields \cite{gdnstrf}.
The narrow resonances can also be used in schemes for obtaining ''slow
light'' \cite{slowlight}.

Pump-probe spectroscopy of Bose-Einstein condensates represents an
interesting application of the ideas presented here. Bragg spectroscopy has
recently been demonstrated in condensates \cite{bragg1}\cite{bragg2} as well
as the stimulated generation of matter waves in a condensate by Rayleigh
scattering \cite{superrrad}. Pump-probe spectroscopy using electronic
excited states would be unfeasible in condensates since there are no stable
trapped condensates with electronic states which may be populated.
Currently, the only multi-component condensates consist of two hyperfine
states in $^{87}Rb$ \cite{Rb} and the Zeeman states in the $F=1$ manifold of
optically trapped $Na$ \cite{Na}. Consequently, pump-probe spectroscopy
would necessarily involve Raman transitions between stable ground states in
the manner proposed here. The RIR spectrum of a weakly interacting Bose
condensate should yield information about the spectrum of elementary
excitations in the condensate which for small momenta have a linear
dispersion relation while, for large momenta, have a quadratic dispersion
similar to that of free atoms, but with a shift due to the mean-field
interactions in the condensate. In addition, the line-widths of the RIR\
spectrum should be given by the zero point motion of the condensate in the
trapping potential provided $\left( \frac{\hbar }{m\Delta x}\right)
k_{p1,2}>\gamma $ and $\Delta x$ is the size of the condensate. However, the
direct application of the results presented here to a condensate would be
erroneous since a correct calculation of the RIR spectrum would have to
account for the mean-field interactions between the atoms. This will be
pursued in future work.

\section{Acknowledgments}

C. P. S. and P. R. B. are pleased to acknowledge helpful discussions with B.
Dubetsky. This research is supported by the National Science Foundation
under Grant No. PHY-9800981 and by the U. S. Army Research Office under
Grant No. DAAG55-97-0113 and No. DAAD19-00-1-0412.

\section{Appendix A - Matrix elements of ${\bf V}_{Id}$}

The interaction with the probe field may be expressed in terms of an
effective two-photon interaction Hamiltonian similar to that in Eq. (\ref
{twophot}), 
\begin{equation}
{\bf V}_{I}=\hbar \sum_{{\bf p}}\left( \frac{\chi _{p}g_{2}^{\ast }}{\Delta }%
e^{-i\Omega _{p}t}\left| 2,{\bf p}+\hbar {\bf k}_{p2}\right\rangle
\left\langle 1,{\bf p}\right| a_{2}^{\dagger }+\frac{\chi _{p}g_{1}^{\ast }}{%
\Delta }e^{-i\Omega _{p}t}\left| 1,{\bf p}+\hbar {\bf k}_{p1}\right\rangle
\left\langle 1,{\bf p}\right| a_{1}^{\dagger }+h.c.\right)  \label{Hp}
\end{equation}
where $\chi _{p}=\frac{-1}{2\hbar }d_{e1}E_{p}$ is the Rabi frequency for
the probe field and $d_{e1}=\left. <e|{\bf d\cdot }\widehat{{\bf \epsilon }}%
|1>\right. $ is a dipole matrix element. The matrix representation of ${\bf V%
}_{I}$ with respect to the bare state basis has the following nonvanishing
elements 
\begin{align}
\left\langle 1,{\bf p};n_{1},n_{2}\right| {\bf V}_{I}\left| 2,{\bf p}+\hbar 
{\bf k}_{12}+\hbar {\bf k}_{p1};n_{1},n_{2}+1\right\rangle & =\hbar
G_{2}e^{i\Omega _{p}t};  \nonumber \\
\left\langle 1,{\bf p};n_{1},n_{2}\right| {\bf V}_{I}\left| 1,{\bf p}+\hbar 
{\bf k}_{p1};n_{1}+1,n_{2}\right\rangle & =\hbar G_{1}e^{i\Omega _{p}t};
\label{Vi} \\
\left\langle 1,{\bf p-}\hbar {\bf k}_{p1};n_{1}-1,n_{2}\right| {\bf V}%
_{I}\left| 2,{\bf p}+\hbar {\bf k}_{12};n_{1}-1,n_{2}+1\right\rangle &
=\hbar G_{2}e^{i\Omega _{p}t};  \nonumber \\
\left\langle 1,{\bf p-}\hbar {\bf k}_{p1};n_{1}-1,n_{2}\right| {\bf V}%
_{I}\left| 1,{\bf p};n_{1},n_{2}\right\rangle & =\hbar G_{1}e^{i\Omega
_{p}t};  \nonumber
\end{align}
and the hermitian conjugates of Eqs. (\ref{Vi}). In Eqs. (\ref{Vi}), $G_{1}$
and $G_{2}$ are two-photon probe Rabi frequencies defined as $G_{1}=\chi
_{p}^{\ast }\chi _{1}/\Delta ;$\ $G_{2}=\chi _{p}^{\ast }\chi _{2}/\Delta .$
Note that the coupling of the 1 and 2 manifolds to manifolds other than the
0 manifold has been ignored.

The matrix ${\bf V}_{Id}$ represents the interaction with the probe field in
the dressed state basis and is defined as 
\begin{equation}
{\bf V}_{Id}={\bf TV}_{I}{\bf T}^{\dagger }{\bf ;}
\end{equation}
where ${\bf T}$ is given by the block diagonal matrix 
\begin{equation}
{\bf T}=diag\left[ {\bf T}_{1}({\bf p}),{\bf T}_{0}({\bf p}),{\bf T}_{2}(%
{\bf p})\right] .  \label{dressed_trans}
\end{equation}
The Rabi frequencies $\chi _{p}$, $\chi _{1}$, and $\chi _{2}$ may be
expressed as 
\begin{equation}
\chi _{p}=|\chi _{p}|e^{i\phi };\ \chi _{1}=|\chi _{1}|e^{i\phi _{1}};\ \chi
_{2}=|\chi _{2}|e^{i\phi _{2}};
\end{equation}
so that $\phi _{d}=\phi _{2}-\phi _{1}+\frac{\pi }{2}(1-\psi )$ since $%
G=|G|e^{i\phi _{d}}$. The matrix elements are:

\begin{mathletters}
\label{abs}
\begin{eqnarray}
&<&A_{1}|{\bf V}_{I}|A_{0}>=\hbar e^{i(\phi -\phi _{1})}e^{-i\Omega
_{p}t}\sin \theta _{0}(-|G_{2}|\cos \theta _{1}+\psi |G_{1}|\sin \theta
_{1});  \label{Vid1} \\
&<&A_{1}|{\bf V}_{I}|B_{0}>=\hbar e^{i(\phi -\phi _{1})}e^{-i\Omega
_{p}t}\cos \theta _{0}(|G_{2}|\cos \theta _{1}-\psi |G_{1}|\sin \theta _{1});
\\
&<&B_{1}|{\bf V}_{I}|A_{0}>=\hbar e^{i(\phi -\phi _{1})}e^{-i\Omega
_{p}t}\sin \theta _{0}(-\psi |G_{1}|\cos \theta _{1}-|G_{2}|\sin \theta
_{1}); \\
&<&B_{1}|{\bf V}_{I}|B_{0}>=\hbar e^{i(\phi -\phi _{1})}e^{-i\Omega
_{p}t}\cos \theta _{0}(\psi |G_{1}|\cos \theta _{1}+|G_{2}|\sin \theta _{1});
\\
&<&A_{0}|{\bf V}_{I}|A_{2}>=\hbar e^{i(\phi -\phi _{1})}e^{-i\Omega
_{p}t}\sin \theta _{2}(-|G_{2}|\cos \theta _{0}+\psi |G_{1}|\sin \theta
_{0}); \\
&<&A_{0}|{\bf V}_{I}|B_{2}>=\hbar e^{i(\phi -\phi _{1})}e^{-i\Omega
_{p}t}\cos \theta _{2}(|G_{2}|\cos \theta _{0}-\psi |G_{1}|\sin \theta _{0});
\\
&<&B_{0}|{\bf V}_{I}|A_{2}>=\hbar e^{i(\phi -\phi _{1})}e^{-i\Omega
_{p}t}\sin \theta _{2}(-\psi |G_{1}|\cos \theta _{0}-|G_{2}|\sin \theta
_{0}); \\
&<&B_{0}|{\bf V}_{I}|B_{2}>=\hbar e^{i(\phi -\phi _{1})}e^{-i\Omega
_{p}t}\cos \theta _{2}(\psi |G_{1}|\cos \theta _{0}+|G_{2}|\sin \theta _{0}).
\label{Vid8}
\end{eqnarray}
The other elements follow from the hermiticity of ${\bf V}_{Id}$.

\section{Appendix B}

In this Appendix the absorption coefficient for the probe field is
calculated, without making the secular approximation. The absorption
coefficient, $\alpha $, and index change, $\Delta n$, arise from the
imaginary and real parts of the macroscopic polarization in the
Maxwell-Bloch equations for the probe field. They are given by the
expressions 
\end{mathletters}
\begin{mathletters}
\label{abs}
\begin{eqnarray}
\alpha &=&\alpha _{0}%
\mathop{\rm Im}%
\left( \frac{\gamma \rho _{1e}^{\prime }}{\chi _{p}^{\ast }}V\right) ;
\label{14a} \\
\Delta n &=&-\alpha _{0}k_{p}^{-1}%
\mathop{\rm Re}%
\left( \frac{\gamma \rho _{1e}^{\prime }}{\chi _{p}^{\ast }}V\right) ;
\label{14b} \\
\alpha _{0} &=&\frac{k_{p}Nd_{1e}^{2}}{2\hbar \epsilon _{0}\gamma }.
\end{eqnarray}
where $V$ is the volume and $\rho _{1e}^{\prime }$ is the part of the bare
state density matrix element $\rho _{1e}({\bf R},t)$ which is proportional
to $e^{-i({\bf k}_{p}\cdot {\bf R}-\Omega _{p}t)}$, which we denote by $\rho
_{1e}^{\prime }({\bf R},t)$.

Before proceeding, we note that in this Appendix all summations over
momentum states have been converted to an integration over a continuum of
states via the standard substitution $\sum_{{\bf p}}\rightarrow \frac{V}{%
(2\pi \hbar )^{3}}\int d^{3}p$. The coefficient $\rho _{1e}^{\prime }$ is
related to the coherence in position space, $\rho _{1e}^{\prime }({\bf R},t)$%
, and the momentum space density matrix elements, $\rho _{1e}({\bf p},{\bf p}%
^{\prime };t)=\rho _{1e}^{^{\prime }}({\bf p},{\bf p}^{\prime };t)e^{i\Omega
_{p}t},$ by 
\end{mathletters}
\begin{eqnarray}
\rho _{1e}^{\prime }({\bf R},t) &=&\rho _{1e}^{\prime }e^{-i({\bf k}%
_{p}\cdot {\bf R}-\Omega _{p}t)}  \label{13} \\
&=&\frac{1}{(2\pi \hbar )^{3}}\int \int d^{3}pd^{3}p^{\prime }\rho
_{1e}^{\prime }({\bf p},{\bf p}^{\prime };t)e^{i\Omega _{p}t}e^{i({\bf p}-%
{\bf p}^{\prime })\cdot {\bf R/\hbar }}\delta ({\bf p}-{\bf p}^{\prime
}+\hbar {\bf k}_{p}).  \label{13b}
\end{eqnarray}
The coherence, $\rho _{1e}^{\prime }({\bf p},{\bf p}^{\prime };t)e^{i\Omega
_{p}t}$, has been written in the Schr\"{o}dinger representation and is
obtained from the density matrix for the atom plus pump fields by tracing
over the number of photons in the pump fields, 
\begin{equation}
\rho _{1e}^{\prime }({\bf p},{\bf p}^{\prime };t)=e^{-i\Omega
_{p}t}e^{-i\omega _{1e}t}\sum_{n_{1},n_{2}}\rho _{1e}^{I}({\bf p}%
,n_{1},n_{2};{\bf p}^{\prime },n_{1},n_{2};t).  \label{trace}
\end{equation}
where $\rho _{1e}^{I}({\bf p},n_{1},n_{2};{\bf p}^{\prime },n_{1}^{\prime
},n_{2}^{\prime };t)$ is in the interaction representation with respect to
the internal energy levels and pump fields. One cannot derive a differential
equation for $\rho _{1e}^{^{\prime }}({\bf p},{\bf p}^{\prime };t)$ starting
from the original Hamiltonian in Eq. (\ref{1}) since the excited state has
been adiabatically eliminated from the effective Hamiltonian. However, by
reintroducing an interaction term 
\[
H_{af}=\hbar \left[ g_{1}\left| e\right\rangle \left\langle 1\right| e^{i%
{\bf k}_{1}\cdot {\bf R}}+g_{2}\left| e\right\rangle \left\langle 2\right|
e^{i{\bf k}_{2}\cdot {\bf R}}+\chi _{p}\left| e\right\rangle \left\langle
1\right| e^{i\left( {\bf k}_{1}\cdot {\bf R-\Omega }_{p}t\right) }\right]
+h.c. 
\]
into the Hamiltonian and writing 
\begin{equation}
\dot{\rho}_{1e}^{\prime }({\bf p},{\bf p}^{\prime };t)=\frac{\partial }{%
\partial t}\left( e^{-i\Omega _{p}t}e^{-i\omega
_{1e}t}\sum_{n_{1},n_{2}}\rho _{1e}^{I}({\bf p},n_{1},n_{2};{\bf p}^{\prime
},n_{1},n_{2};t)\right) .
\end{equation}
one finds the equation of motion for $\rho _{1e}^{I}({\bf p},n_{1},n_{2};%
{\bf p}^{\prime },n_{1}^{\prime },n_{2}^{\prime })$ to be 
\begin{eqnarray}
\dot{\rho}_{1e}^{I}({\bf p},n_{1},n_{2};{\bf p}^{\prime },n_{1}^{\prime
},n_{2}^{\prime }) &=&\left[ -i\omega _{{\bf pp}^{\prime }}-(\gamma +\gamma
_{e})/2\right] \rho _{1e}^{I}({\bf p},n_{1},n_{2};{\bf p}^{\prime
},n_{1}^{\prime },n_{2}^{\prime })  \nonumber \\
&&-i\chi _{1}^{\ast }e^{i(\Omega _{1}-\omega _{e1})t}\left[ \rho _{ee}^{I}(%
{\bf p}+\hbar {\bf k}_{1},n_{1}-1,n_{2};{\bf p}^{\prime },n_{1}^{\prime
},n_{2}^{\prime })\right.  \nonumber \\
&&\left. -\rho _{11}^{I}({\bf p},n_{1},n_{2};{\bf p}^{\prime }-\hbar {\bf k}%
_{1},n_{1}^{\prime }+1,n_{2}^{\prime })\right]  \nonumber \\
&&-i\chi _{p}^{\ast }e^{i(\Omega _{p}-\omega _{e1})t}\left[ \rho _{ee}^{I}(%
{\bf p}+\hbar {\bf k}_{p},n_{1},n_{2};{\bf p}^{\prime },n_{1}^{\prime
},n_{2}^{\prime })\right.  \nonumber \\
&&\left. -\rho _{11}^{I}({\bf p},n_{1},n_{2};{\bf p}^{\prime }-\hbar {\bf k}%
_{p},n_{1}^{\prime },n_{2}^{\prime })\right]  \nonumber \\
&&+i\chi _{2}^{\ast }e^{i(\Omega _{2}+\omega _{2e})t}\rho _{12}^{I}({\bf p}%
,n_{1},n_{2};{\bf p}^{\prime }-\hbar {\bf k}_{2},n_{1}^{\prime
},n_{2}^{\prime }+1),
\end{eqnarray}
where the $t$ argument has been dropped. By carrying out the trace in Eq. (%
\ref{trace}), one obtains terms such as, 
\[
\rho _{11}^{I}({\bf p};{\bf p}^{\prime }-\hbar {\bf k}_{1})=%
\sum_{n_{1},n_{2}}\rho _{11}^{I}({\bf p},n_{1},n_{2};{\bf p}^{\prime }-\hbar 
{\bf k}_{1},n_{1}+1,n_{2}), 
\]
so that the equation of motion for $\rho _{1e}^{\prime }({\bf p};{\bf p}%
^{\prime })$ has a form which is identical to that which would have been
obtained using classical pump fields, 
\begin{eqnarray}
\dot{\rho}_{1e}^{^{\prime }}({\bf p};{\bf p}^{\prime }) &=&-\left[ i(\Omega
_{p}+\omega _{1e}+\omega _{{\bf pp}^{\prime }})+(\gamma +\gamma _{e})/2%
\right] \rho _{1e}^{^{\prime }}({\bf p};{\bf p}^{\prime })  \nonumber \\
&&-i\chi _{p}^{\ast }\left[ \rho _{ee}^{I}({\bf p}+\hbar {\bf k}_{p};{\bf p}%
^{\prime })-\rho _{11}^{I}({\bf p};{\bf p}^{\prime }-\hbar {\bf k}_{p})%
\right]  \nonumber \\
&&-i\chi _{1}^{\ast }e^{-i\delta _{p1}t}\left[ \rho _{ee}^{I}({\bf p}+\hbar 
{\bf k}_{1};{\bf p}^{\prime })-\rho _{11}^{I}({\bf p};{\bf p}^{\prime
}-\hbar {\bf k}_{1})\right]  \nonumber \\
&&+i\chi _{2}^{\ast }e^{-i\delta _{p2}t}\rho _{12}^{I}({\bf p;p}^{\prime
}-\hbar {\bf k}_{2}).  \label{densI}
\end{eqnarray}
In terms of the perturbation series solution 
\begin{mathletters}
\label{abs}
\begin{eqnarray}
\rho _{11}^{I}({\bf p};{\bf p}^{\prime }) &=&\rho _{11}^{(0)}({\bf p};{\bf p}%
^{\prime })+\rho _{11}^{+}({\bf p};{\bf p}^{\prime })e^{i\delta _{p1}t}+\rho
_{11}^{-}({\bf p};{\bf p}^{\prime })e^{-i\delta _{p1}t};  \label{pert1} \\
\rho _{12}^{I}({\bf p};{\bf p}^{\prime }) &=&\left[ \rho _{12}^{(0)}({\bf p};%
{\bf p}^{\prime })+\rho _{12}^{+}({\bf p};{\bf p}^{\prime })e^{i\delta
_{p1}t}+\rho _{12}^{-}({\bf p};{\bf p}^{\prime })e^{-i\delta _{p1}t}\right]
e^{i\delta _{12}t};  \label{pert2}
\end{eqnarray}
where $\rho _{jj^{\prime }}^{(0)}({\bf p};{\bf p}^{\prime })$ are
independent of $\chi _{p}$ and $\rho _{jj^{\prime }}^{\pm }({\bf p};{\bf p}%
^{\prime })$ are {\it linear} in $\chi _{p}$, the steady state solution for
large detuning is 
\end{mathletters}
\begin{equation}
\rho _{1e}^{\prime }({\bf p};{\bf p}^{\prime })\approx \frac{1}{\Delta }%
\left( \chi _{2}^{\ast }\rho _{12}^{+}({\bf p;p}^{\prime }-\hbar {\bf k}%
_{2})+\chi _{1}^{\ast }\rho _{11}^{+}({\bf p};{\bf p}^{\prime }-\hbar {\bf k}%
_{1})+\chi _{p}^{\ast }\rho _{11}^{(0)}({\bf p};{\bf p}^{\prime }-\hbar {\bf %
k}_{p})\right) .
\end{equation}
By making a change of variables, ${\bf p}-{\bf p}^{\prime }\rightarrow {\bf p%
}-{\bf p}^{\prime }-\hbar {\bf k}_{p}$ in Eq. (\ref{13b}), one gets, 
\begin{eqnarray}
\rho _{1e}^{\prime }({\bf R},t) &=&\frac{e^{-i({\bf k}_{p}\cdot {\bf R}%
-\Omega _{p}t)}}{(2\pi \hbar )^{3}\Delta }\int \int d^{3}pd^{3}p^{\prime
}e^{i({\bf p}-{\bf p}^{^{\prime }})\cdot {\bf R/\hbar }}\delta ({\bf p}-{\bf %
p}^{\prime })  \nonumber \\
&&\times \{\chi _{2}^{\ast }\left( \rho _{12}^{+}({\bf p;p}^{\prime }+\hbar 
{\bf k}_{p2})+\rho _{12}^{+}({\bf p-}\hbar {\bf k}_{p1}{\bf ;p}^{\prime
}+\hbar {\bf k}_{12})\right) +  \nonumber \\
&&\chi _{1}^{\ast }\left( \rho _{11}^{+}({\bf p};{\bf p}^{\prime }+\hbar 
{\bf k}_{p1})+\rho _{11}^{+}({\bf p-}\hbar {\bf k}_{p1};{\bf p}^{\prime
})\right) +\chi _{p}^{\ast }\rho _{11}^{(0)}({\bf p};{\bf p}^{\prime })\}
\label{15}
\end{eqnarray}
One must now obtain equations for $\rho _{11}^{I}({\bf p};{\bf p}^{\prime })$
and $\rho _{12}^{I}({\bf p};{\bf p}^{\prime })$ using the effective
Hamiltonian ${\bf H}_{tot}{\bf =H+V}_{I}$, solve these equations, and then
extract $\rho _{12}^{+}({\bf p};{\bf p}^{\prime })$ and $\rho _{11}^{+}({\bf %
p};{\bf p}^{\prime })$ from these solutions using Eqs. (\ref{pert1}-\ref
{pert2}).

The terms appearing in Eq. (\ref{15}) may be expressed in terms of the
dressed state density matrix elements. If one is interested only in terms
linear in $\chi _{p}$, one can expand the dressed state density matrix to
first order in $\chi _{p}$ as 
\begin{mathletters}
\label{abs}
\begin{eqnarray}
\rho _{II^{\prime }} &=&\left[ \rho _{II^{\prime }}^{(0)}+\rho _{II^{\prime
}}^{+}e^{i\left( \Omega _{p}-\omega _{10}\right) t}+\rho _{II^{\prime
}}^{-}e^{-i\left( \Omega _{p}-\omega _{10}\right) t}\right] e^{i\omega
_{10}t};  \label{dresspert} \\
\rho _{JJ^{\prime }} &=&\left[ \rho _{JJ^{\prime }}^{(0)}+\rho _{JJ^{\prime
}}^{+}e^{i\left( \Omega _{p}+\omega _{20}\right) t}+\rho _{JJ^{\prime
}}^{-}e^{-i\left( \Omega _{p}+\omega _{20}\right) t}\right] e^{i\omega
_{20}t};  \label{dresspert2}
\end{eqnarray}
where $I,J=\{A_{0},B_{0}\}$, $I^{\prime }=\{A_{1},B_{1}\},$ and $J^{\prime
}=\{A_{2},B_{2}\}$. The $\rho _{ij}^{+}({\bf p};{\bf p}^{\prime })$ needed
in Eq.(\ref{15}) can be expressed in terms of dressed state density matrix
elements using Eqs. (\ref{bare2dressed},\ref{T},\ref{dens}) as 
\end{mathletters}
\begin{mathletters}
\label{abs}
\begin{eqnarray}
\rho _{11}^{+}({\bf p};{\bf p}+\hbar {\bf k}_{p1}) &=&\sin \theta _{0}({\bf p%
})\sin \theta _{1}({\bf p})\rho _{A_{0}A_{1}}^{+}-\sin \theta _{0}({\bf p}%
)\cos \theta _{1}({\bf p})\rho _{A_{0}B_{1}}^{+}  \nonumber \\
&&-\cos \theta _{0}({\bf p})\sin \theta _{1}({\bf p})\rho
_{B_{0}A_{1}}^{+}+\cos \theta _{0}({\bf p})\cos \theta _{1}({\bf p})\rho
_{B_{0}B_{1}}^{+};  \label{17a} \\
\rho _{11}^{+}({\bf p-}\hbar {\bf k}_{p1};{\bf p}) &=&\sin \theta _{2}({\bf p%
})\sin \theta _{0}({\bf p})\rho _{A_{2}A_{0}}^{+}-\sin \theta _{2}({\bf p}%
)\cos \theta _{0}({\bf p})\rho _{A_{2}B_{0}}^{+}  \nonumber \\
&&-\cos \theta _{2}({\bf p})\sin \theta _{0}({\bf p})\rho
_{B_{2}A_{0}}^{+}+\cos \theta _{2}({\bf p})\cos \theta _{0}({\bf p})\rho
_{B_{2}B_{0}}^{+}; \\
\rho _{12}^{+}({\bf p;p}+\hbar {\bf k}_{p2}) &=&e^{i\phi _{d}}\left( -\sin
\theta _{0}({\bf p})\cos \theta _{1}({\bf p})\rho _{A_{0}A_{1}}^{+}-\sin
\theta _{0}({\bf p})\sin \theta _{1}({\bf p})\rho _{A_{0}B_{1}}^{+}\right.  
\nonumber \\
&&\left. +\cos \theta _{0}({\bf p})\cos \theta _{1}({\bf p})\rho
_{B_{0}A_{1}}^{+}+\cos \theta _{0}({\bf p})\sin \theta _{1}({\bf p})\rho
_{B_{0}B_{1}}^{+}\right) ; \\
\rho _{12}^{+}({\bf p-}\hbar {\bf k}_{p1}{\bf ;p}+\hbar {\bf k}_{12})
&=&e^{i\phi _{d}}\left( -\sin \theta _{2}({\bf p})\cos \theta _{0}({\bf p}%
)\rho _{A_{2}A_{0}}^{+}-\sin \theta _{2}({\bf p})\sin \theta _{0}({\bf p}%
)\rho _{A_{2}B_{0}}^{+}\right.   \nonumber \\
&&\left. +\cos \theta _{2}({\bf p})\cos \theta _{0}({\bf p})\rho
_{B_{2}A_{0}}^{+}+\cos \theta _{2}({\bf p})\sin \theta _{0}({\bf p})\rho
_{B_{2}B_{0}}^{+}\right) .  \label{17d}
\end{eqnarray}

The state vector in the Schr\"{o}dinger representation may be expanded in
terms of the dressed states for the 0, 1, and 2 manifolds, 
\end{mathletters}
\[
|\Psi >=c_{A_{0}}^{s}({\bf p)}|A_{0}>+c_{B_{0}}^{s}({\bf p)}%
|B_{0}>+c_{A_{1}}^{s}({\bf p)}|A_{1}>+c_{B_{1}}^{s}({\bf p)}%
|B_{1}>+c_{A_{2}}^{s}({\bf p)}|A_{2}>+c_{B_{2}}^{s}({\bf p)}|B_{2}>. 
\]
In the following, the momentum labels are suppressed. The Schr\"{o}dinger
equation for the dressed state amplitudes is then given by 
\begin{equation}
i\hbar {\bf \dot{c}}=\left( {\bf H}_{o}+{\bf V}_{Id}\right) {\bf c}
\label{dressschro}
\end{equation}
where 
\begin{equation}
{\bf H}%
_{o}=diag(E_{A}^{(1)},E_{B}^{(1)},E_{A}^{(0)},E_{B}^{(0)},E_{A}^{(2)},E_{B}^{(2)})
\end{equation}
and matrix elements of ${\bf V}_{Id}$ are given in Eqs. (\ref{Vid1}-\ref
{Vid8}).

Using Eq. (\ref{dressschro}) along with the ground state decay rate $\gamma $
and incoherent pumping to the 0 manifold, one finds that density matrix
elements for the six dressed states in the three manifolds (0), (1), (2),
evolve as 
\begin{equation}
\left( \frac{d}{dt}+\gamma \right) {\bf \rho }_{d}=\frac{1}{i\hbar }[{\bf H}%
_{o}+{\bf V}_{Id},{\bf \rho }_{d}]+{\bf \Lambda }_{D};  \label{bloch}
\end{equation}
where 
\begin{equation}
{\bf \rho }_{d}({\bf p},{\bf p}^{\prime })={\bf c(p)c}^{\dagger }{\bf %
(p^{\prime })}=\left( 
\begin{array}{cccccc}
\rho _{A_{1}A_{1}} & \rho _{A_{1}B_{1}} & \rho _{A_{1}A_{0}} & \rho
_{A_{1}B_{o}} & \rho _{A_{1}A_{2}} & \rho _{A_{1}B_{2}} \\ 
\rho _{B_{1}A_{1}} & \rho _{B_{1}B_{1}} & \rho _{B_{1}A_{0}} & \rho
_{B_{1}B_{o}} & \rho _{B_{1}A_{2}} & \rho _{B_{1}B_{2}} \\ 
\rho _{A_{0}A_{1}} & \rho _{A_{0}B_{1}} & \rho _{A_{0}A_{0}} & \rho
_{A_{0}B_{o}} & \rho _{A_{0}A_{2}} & \rho _{A_{0}B_{2}} \\ 
\rho _{B_{0}A_{1}} & \rho _{B_{0}B_{1}} & \rho _{B_{0}A_{0}} & \rho
_{B_{0}B_{o}} & \rho _{B_{0}A_{2}} & \rho _{B_{0}B_{2}} \\ 
\rho _{A_{2}A_{1}} & \rho _{A_{2}B_{1}} & \rho _{A_{2}A_{0}} & \rho
_{A_{2}B_{o}} & \rho _{A_{2}A_{2}} & \rho _{A_{2}B_{2}} \\ 
\rho _{B_{2}A_{1}} & \rho _{B_{2}B_{1}} & \rho _{B_{2}A_{0}} & \rho
_{B_{2}B_{o}} & \rho _{B_{2}A_{2}} & \rho _{B_{2}B_{2}}
\end{array}
\right)  \label{dens}
\end{equation}
and the pumping matrix, ${\bf \Lambda }_{d},$ has the block diagonal form 
\begin{equation}
{\bf \Lambda }_{D}=diag({\bf 0,\Lambda }_{d}({\bf p,p}^{\prime }{\bf ),0}),
\end{equation}
where ${\bf \Lambda }_{d}({\bf p)}$ has the basic structure given in Eq. (%
\ref{pump}), modified to allow for ${\bf p,p}^{\prime }$ coherence. In
particular, the off diagonal elements of ${\bf \Lambda }_{d}({\bf p,p}%
^{\prime }{\bf )}$ that give rise to the nonsecular contribution to the line
shape are of the form 
\[
\Lambda _{AB}({\bf p,p}^{\prime }+\hbar {\bf k}_{12}{\bf )=}\Lambda _{2}(%
{\bf p}+\hbar {\bf k}_{12},{\bf p}^{\prime }+\hbar {\bf k}_{12})\cos \left[
\theta _{0}({\bf p)}\right] \sin \left[ \theta _{0}({\bf p}^{\prime }{\bf )}%
\right] -\Lambda _{1}({\bf p,p}^{\prime })\sin \left[ \theta _{0}({\bf p)}%
\right] \cos \left[ \theta _{0}({\bf p}^{\prime }{\bf )}\right] . 
\]

The general form of the solution is linked to the incoherent pumping of
levels 1 and 2. For a subrecoil cooled vapor, the pumping rate density for
bare state density matrix elements $\rho _{ij}({\bf p};{\bf p}^{\prime })$
is assumed to be 
\begin{equation}
\Lambda _{ij}({\bf p},{\bf p}^{\prime })=\Lambda _{i}V^{-1}(2\pi \hbar
)^{3}\delta ({\bf p})\delta ({\bf p}-{\bf p}^{\prime })\delta _{ij}
\label{pump2}
\end{equation}
where $\Lambda _{i}V^{-1}$ has the dimensions of $($volume$\times $time$%
)^{-1}$ and can be interpreted as the pumping rate to state 1 or 2 in
position space. With this form of pumping, $\rho _{ij}^{I}({\bf p};{\bf p}%
^{\prime }),$ must be proportional to $\delta ({\bf p}-{\bf p}^{\prime
}-\hbar {\bf k}^{\prime })$ where ${\bf k}^{\prime }$ is some algebraic
combination of the pump and probe field propagation vectors. To obtain $\rho
_{1e}^{\prime }$ from Eq. (\ref{15}), one must keep only those terms in the
integrand of Eq. (\ref{15}) proportional to $\delta \left( {\bf p}-{\bf p}%
^{\prime }\right) .$ The $\rho _{11}^{(0)}({\bf p};{\bf p})$ term in Eq. (%
\ref{15}) makes no contribution to the absorption since it is real and will
be ignored from this point on.

The incoherent pumping of states 1 and 2 populate two {\it different }%
manifolds. The pumping of state 1 populates the $\left( {\bf p}%
=0,n_{1},n_{2}\right) $ manifold while the pumping of state 2 populates the $%
\left( {\bf p}=-\hbar {\bf k}_{12},n_{1},n_{2}\right) $ manifold since this
manifold involves state 2 with zero momentum. Thus in viewing absorption or
emission, two {\it distinct }initial state manifolds must be included,
leading to the possibility of sixteen rather than eight components of the
spectrum. Here we set $\Lambda _{2}=0.$

Substituting Eqs. (\ref{17a}-\ref{17d}) into Eq. (\ref{15}) and using the
steady state solutions for $\rho _{II^{\prime }}^{+}$ and $\rho _{JJ^{\prime
}}^{+}$ obtained from Eqs. (\ref{dresspert}-\ref{dresspert2}) and Eq. (\ref
{bloch}), one finds, after some manipulation, the final expression for the
absorption coefficient, 
\begin{mathletters}
\label{abs}
\begin{eqnarray}
\left( \frac{\alpha }{\alpha _{0}}\right) &=&\left( \frac{\alpha }{\alpha
_{0}}\right) _{\sec }+\left( \frac{\alpha }{\alpha _{0}}\right) _{ns};
\label{absorb} \\
\left( \frac{\alpha }{\alpha _{0}}\right) _{ns} &=&\frac{|G|}{|\Delta |}%
\frac{\Lambda _{AB}}{\gamma }\frac{\sin 2\theta _{0}}{2\left( (\omega
_{AB}^{(0)})^{2}+\gamma ^{2}\right) }\left[ -\left( \psi \eta \sin \theta
_{1}-\frac{1}{\eta }\cos \theta _{1}\right) ^{2}\left( \Gamma
_{A_{0}A_{1}}(\Delta ^{\prime })+\Gamma _{B_{0}A_{1}}(\Delta ^{\prime
})\right) \right.  \nonumber \\
&&-\left( \psi \eta \cos \theta _{1}+\frac{1}{\eta }\sin \theta _{1}\right)
^{2}\left( \Gamma _{A_{0}B_{1}}(\Delta ^{\prime })+\Gamma
_{B_{0}B_{1}}(\Delta ^{\prime })\right)  \nonumber \\
&&+\left. \left( \eta ^{2}-\eta ^{-2}-2\psi \cot 2\theta _{0}\right) \left(
\sin ^{2}\theta _{2}\left( \Gamma _{A_{0}A_{2}}(\Delta ^{\prime })+\Gamma
_{B_{0}A_{2}}(\Delta ^{\prime })\right) +\cos ^{2}\theta _{2}\left( \Gamma
_{A_{0}B_{2}}(\Delta ^{\prime })+\Gamma _{B_{0}B_{2}}(\Delta ^{\prime
})\right) \right) \right] ; \\
\Gamma _{A_{0}A_{1}}(\Delta ^{\prime }) &=&\left[ \left( \Delta ^{\prime
}-\omega _{\hbar {\bf k}_{p1}}-\frac{1}{2}(\omega _{AB}^{(0)}-\omega
_{AB}^{(1)})\right) \omega _{AB}^{(0)}+\gamma ^{2}\right] L_{A_{0}A_{1}}(%
\Delta ^{\prime }); \\
\Gamma _{B_{0}A_{1}}(\Delta ^{\prime }) &=&\left[ -\left( \Delta ^{\prime
}-\omega _{\hbar {\bf k}_{p1}}+\frac{1}{2}(\omega _{AB}^{(0)}+\omega
_{AB}^{(1)})\right) \omega _{AB}^{(0)}+\gamma ^{2}\right] L_{B_{0}A_{1}}(%
\Delta ^{\prime }); \\
\Gamma _{A_{0}B_{1}}(\Delta ^{\prime }) &=&\left[ \left( \Delta ^{\prime
}-\omega _{\hbar {\bf k}_{p1}}-\frac{1}{2}(\omega _{AB}^{(0)}+\omega
_{AB}^{(1)})\right) \omega _{AB}^{(0)}+\gamma ^{2}\right] L_{A_{0}B_{1}}(%
\Delta ^{\prime }); \\
\Gamma _{B_{0}B_{1}}(\Delta ^{\prime }) &=&\left[ -\left( \Delta ^{\prime
}-\omega _{\hbar {\bf k}_{p1}}+\frac{1}{2}(\omega _{AB}^{(0)}-\omega
_{AB}^{(1)})\right) \omega _{AB}^{(0)}+\gamma ^{2}\right] L_{B_{0}B_{1}}(%
\Delta ^{\prime }); \\
\Gamma _{A_{0}A_{2}}(\Delta ^{\prime }) &=&\left[ -\left( \Delta ^{\prime
}+\omega _{\hbar {\bf k}_{p1}}+\frac{1}{2}(\omega _{AB}^{(0)}-\omega
_{AB}^{(2)})\right) \omega _{AB}^{(0)}+\gamma ^{2}\right] L_{A_{0}A_{2}}(%
\Delta ^{\prime }); \\
\Gamma _{B_{0}A_{2}}(\Delta ^{\prime }) &=&\left[ \left( \Delta ^{\prime
}+\omega _{\hbar {\bf k}_{p1}}-\frac{1}{2}(\omega _{AB}^{(0)}+\omega
_{AB}^{(2)})\right) \omega _{AB}^{(0)}+\gamma ^{2}\right] L_{B_{0}A_{2}}(%
\Delta ^{\prime }); \\
\Gamma _{A_{0}B_{2}}(\Delta ^{\prime }) &=&\left[ -\left( \Delta ^{\prime
}+\omega _{\hbar {\bf k}_{p1}}+\frac{1}{2}(\omega _{AB}^{(0)}+\omega
_{AB}^{(2)})\right) \omega _{AB}^{(0)}+\gamma ^{2}\right] L_{A_{0}B_{2}}(%
\Delta ^{\prime }); \\
\Gamma _{B_{0}B_{2}}(\Delta ^{\prime }) &=&\left[ \left( \Delta ^{\prime
}+\omega _{\hbar {\bf k}_{p1}}-\frac{1}{2}(\omega _{AB}^{(0)}-\omega
_{AB}^{(2)})\right) \omega _{AB}^{(0)}+\gamma ^{2}\right] L_{B_{0}B_{2}}(%
\Delta ^{\prime });  \label{ablast}
\end{eqnarray}
where $\Lambda _{AB}=-\frac{1}{2}\Lambda _{1}\sin \left( 2\theta _{0}\right) 
$ and $\left( \frac{\alpha }{\alpha _{0}}\right) _{\sec }$ is given by Eq. (%
\ref{absorb2}). The reason the absorption coefficient is expressible as a
sum of the secular term plus a non-secular term is linked to the fact that
the secular approximation consists solely of neglecting the off-diagonal
components of ${\bf \Lambda }_{d}$. Since the first order solutions, $\rho
_{II^{\prime }}^{+}$ and $\rho _{JJ^{\prime }}^{+}$, are linear in the
pumping terms, $\left( \frac{\alpha }{\alpha _{0}}\right) _{\sec }$ contain
terms proportional to $\Lambda _{A}$ and $\Lambda _{B}$ while $\left( \frac{%
\alpha }{\alpha _{0}}\right) _{ns}$ is proportional to $\Lambda _{AB}$. This
simplification would not occur for a more complex decay scheme for states $%
|1>$ and $|2>$ since the decay would couple density matrix elements in a
field dependent manner (see \cite{semiclassical}).

The nonsecular term, $\left( \frac{\alpha }{\alpha _{0}}\right) _{ns}$,
consist of dispersion-like structures centered at the same locations as the
resonances in $\left( \frac{\alpha }{\alpha _{0}}\right) _{\sec }.$ In the
secular limit, $\left( \frac{\alpha }{\alpha _{0}}\right) _{ns}\ll \left( 
\frac{\alpha }{\alpha _{0}}\right) _{\sec }$ and $\left( \frac{\alpha }{%
\alpha _{0}}\right) _{ns}$ can usually be ignored. Notice that if one
chooses $\psi $ and $\eta $ such that a pair of absorption lines in $\left( 
\frac{\alpha }{\alpha _{0}}\right) _{\sec }$ vanish, then the corresponding
terms in $\left( \frac{\alpha }{\alpha _{0}}\right) _{ns}$ also vanish so
that $\left( \frac{\alpha }{\alpha _{0}}\right) $ is identically zero.
However, this will not be true for the gain terms in $\left( \frac{\alpha }{%
\alpha _{0}}\right) _{\sec }$ since the corresponding terms in $\left( \frac{%
\alpha }{\alpha _{0}}\right) _{ns}$ have a different interference
coefficient, $\left( \eta ^{2}-\eta ^{-2}-2\psi \cot 2\theta _{0}\right) $.
Figure 6 shows a plot of the non-secular absorption coefficient for the same
parameters as Fig. 3. In this plot, the non-secular terms are $\sim 1000$
times smaller than the secular terms.

Figure 1. Schematic diagram of atom-field system.

Figure 2. (a) Transitions between the 0 manifold and the 1 and 2 manifolds
leading to probe gain or absorption in the bare state picture. The states in
the 1 and 2 manifolds are displaced from the states in the 0 manifold by an
amount $\sim \hbar \Omega _{1}.$ The Rabi frequencies shown are those which
couple the states from the 0 manifold to the 1 and 2 manifolds. (b) Energy
levels in the dressed state basis. The center of the 1 manifold has an
energy $\hbar \omega _{10}$ above the center of the 0 manifold and
similarly, the center of the 2 manifold is $\hbar \omega _{20}$ below the 0
manifold.(c) Illustration of the coupling of the pump and probe fields to
the dressed states for the $|A_{0}>\rightarrow |A_{1}>$ transition. The Rabi
frequencies shown are those that couple the bare state $|1>$ component of $%
|A_{0}>$ to the bare state components of $|A_{1}>$.

Figure 3. Plot of $\left( \frac{\alpha }{\alpha _{0}}\right) _{\sec }$ for $%
\Lambda _{2}=0$, $\Lambda _{1}/\gamma =1,$ $\psi =-1$ and $\eta =2$ showing
all eight absorption and emission lines. The detuning is $\delta
_{12}/\gamma =300$ and the two-photon pump Rabi frequency is $|G|/\gamma
=250.$ The recoil energies are $\omega _{\hbar {\bf k}_{12}}/\gamma =40$, $%
\omega _{\hbar {\bf k}_{p1}}/\gamma =60$, and $\hbar {\bf k}_{p1}\cdot {\bf k%
}_{12}/M\gamma =80$. Note that $\gamma \ll \gamma _{e}$ where $\gamma _{e}$
is the excited state decay rate.

Figure 4. Plot of $\left( \frac{\alpha }{\alpha _{0}}\right) _{\sec }$
showing destructive interference. Parameters are the same as Fig. 4 except
for $\psi $ and $\eta .$ The solid line corresponds to $\psi =-1$ and $\eta =%
\sqrt{\tan \theta _{1}}=0.8383$ while the dotted line corresponds to $\psi
=+1$ and $\eta =\sqrt{\cot \theta _{1}}=1.1928.$

Figure 5. Plot showing sixteen lines in the $\left( \frac{\alpha }{\alpha
_{0}}\right) _{\sec }$ probe spectrum for $\Lambda _{2}/\gamma =\Lambda
_{1}/\gamma =1$, $\psi =+1$ and $\eta =0.1$. The detuning is $\delta
_{12}/\gamma =500$ and the two-photon pump Rabi frequency is $|G|/\gamma
=750.$ The recoil energies are $\omega _{\hbar {\bf k}_{12}}/\gamma =50$, $%
\omega _{\hbar {\bf k}_{p1}}/\gamma =75$, and $\hbar {\bf k}_{p1}\cdot {\bf k%
}_{12}/M\gamma =-50$.

Figure 6. Plot of $\left( \frac{\alpha }{\alpha _{0}}\right) _{ns}$ for the
same parameters as Fig. (3). The non-secular absorption coefficient has
dispersionlike structures at the same location is the line centers of $%
\left( \frac{\alpha }{\alpha _{0}}\right) _{\sec }.$ The amplitudes of these
non-secular terms is typically $\sim 1000$ times smaller than the secular
line strengths, consistent with $\gamma /\omega _{AB}^{(0)}=0.00177.$
\end{mathletters}

\end{document}